\documentstyle[prl,aps,multicol,epsf]{revtex}
\begin{document}

\title{Stability of the compressible quantum Hall state around the 
half-filled Landau level}
\author{
Kenzo Ishikawa, Nobuki Maeda, and Tetsuyuki Ochiai}
\address{
Department of Physics, Hokkaido University, Sapporo 060-0810, Japan}
\date{\today}
\maketitle
\begin{abstract}
We study the compressible states in the quantum Hall 
system using a mean field theory on the von Neumann lattice. 
In the lowest Landau level, a kinetic energy is generated dynamically 
from Coulomb interaction. 
The compressibility of the state is calculated 
as a function of the filling factor $\nu$ and 
the width $d$ of the spacer between the charge 
carrier layer and dopants. 
The compressibility becomes negative below a critical value of $d$ and 
the state becomes unstable at $\nu=1/2$. 
Within a finite range around $\nu=1/2$, the stable compressible state 
exists above the critical value of $d$. 
\end{abstract}
\draft
\begin{multicols}{2}

The fractional quantum Hall effect (FQHE)\cite{a} is observed 
in a two-dimensional electron system in a strong magnetic field 
around the rational filling factor $\nu$ with an odd denominator. 
The formation of the incompressible liquid state\cite{aa} successfully 
explains the FQHE. 
Recent experiments of the quantum Hall system around the rational filling 
factor $\nu$ with 
an even denominator (ex. 1/2) indicated the formation 
of the Fermi surface\cite{b,c,d,e}. 
The composite fermion theory\cite{f,g} can reasonably explain 
these experiments. 
However, the existence of the stable compressible Fermi liquid at $\nu=1/2$ 
is assumed {\it a priori} in this theory. 
It is more desirable to start from a microscopic theory 
and study the stability of the Fermi liquid state 
in the lowest Landau level space. 
Along this line a microscopic theory of the composite fermion has 
developed\cite{h,i} and an effort is made to obtain the finite 
compressibility at $\nu=1/2$\cite{j}. 
Until now, understanding of the Fermi liquid property of the quantum Hall 
system in the microscopic level is unsatisfactory. 
The purpose of the present work is to study the stability 
of the compressible liquid state around $\nu=1/2$ 
using a mean field theory in the von Neumann lattice formalism\cite{k}. 
We show that the stable Fermi-liquid-like state is formed if a distance 
between the 2d electron layer and dopants is larger than a critical value. 
We call this state the quantum Hall gas (QHG) state. 

In a strong magnetic field the free energy is quenched and kinetic energy, 
pressure, and compressibility are generated from the Coulomb interaction. 
Hence the pressure and compressibility of the QHG 
become negative in the jelium model where the uniform 
positive charge distribution cancels the Coulomb energy 
between electrons\cite{l}. 
In realistic high-mobility samples, however, the positive charges (dopants) 
are separated from the negative charges (carriers) by the spacer in 
order to obtain high-mobility and this separation $d$ gives a positive 
contribution to the energy of the system. 
Owing to this contribution, the pressure and compressibility can become 
positive for a large $d$ and the QHG state becomes stable. 
The critical values $d_{c1}$ and $d_{c2}$ for the pressure and 
compressibility are obtained in the present paper. 

In the lowest Landau level space, 
the total Hamiltonian for a two-dimensional system of this situation 
is given by, 
\begin{eqnarray}
H&=&{1\over2}\int d^2rd^2r':(\rho({\bf r})-\rho_0)
V({\bf r}-{\bf r}')(\rho({\bf r}')-\rho_0):
\nonumber\\
&&+\int d^2rd^2r'\rho({\bf r})(V({\bf r}-{\bf r}')-V_d({\bf r}-{\bf r}'))
\rho_0.
\label{int}
\end{eqnarray}
Here $\rho({\bf r})=\psi^\dagger({\bf r})\psi({\bf r})$ and the electron 
field operator $\psi({\bf r})$ is expanded by the Wannier basis on the 
von Neumann lattice\cite{k} as
\begin{equation}
\psi({\bf r})=\sum_{\bf X}b({\bf X})\langle{\bf r}\vert f_0\otimes\beta_
{\bf X}\rangle,
\end{equation}
where $b$ is the anti-commuting annihilation 
operator and $\bf X$ is an integer-valued two-dimensional coordinate. 
The base functions  $\vert f_0\otimes\beta_{\bf X}\rangle$ are orthonormal 
complete set in the lowest Landau level space. 
Expectation values of the position of the base functions are located at 
two-dimensional lattice sites $a(rm,n/r)$ for ${\bf X}=(m,n)$ 
where $a=\sqrt{2\pi\hbar/{\rm e}B}$. 
The parameter $r$ is fixed later by the minimum energy condition 
at $\nu=1/2$. 
$\rho_0$ is the positive background charge uniformly distributed in the sheet 
separated from the electron gas by a distance $d$. 
The colons in Eq.~(\ref{int}) mean that the normal ordering should be taken. 
We treat the electron as a spinless fermion. 
The potential $V$ and $V_d$ are given by
\begin{equation}
V({\bf r})=q^2/{\rm r},\ 
V_d({\bf r})=q^2/\sqrt{{\rm r}^2+d^2}.
\end{equation}
The second term of the right-hand side in Eq.~(\ref{int}) gives 
$2\pi q^2\nu^2 d/a^2$ per area under the charge neutrality condition. 
For simplicity we set $a=1$ and $\hbar=c=1$. 

By carrying out the integration in Eq.~(\ref{int}), the Hamiltonian 
becomes
\begin{eqnarray} 
H&=&H_0+2\pi q^2\nu^2 L^2 d,\\
H_0&=&{1\over2}\sum_X 
:b^\dagger({\bf X}_1)b({\bf X}'_1)V({\bf X},{\bf Y},{\bf Z})
b^\dagger({\bf X}_2)b({\bf X}'_2):,
\end{eqnarray}
where ${\bf X}={\bf X}_1-{\bf X}'_1$, ${\bf Y}={\bf X}_2-{\bf X}'_2$, 
${\bf Z}={\bf X}_1-{\bf X}'_2$ and $L^2$ is an area of the system. 
The system is, then, translationally invariant on the lattice. 
Therefore, if this invariance is unbroken in the ground state, 
the formation of the Fermi surface is expected. 
Interaction potential $V({\bf X},{\bf Y},{\bf Z})$ is given by
\begin{eqnarray}
\int^{\infty}_{-\infty}{d^2k\over(2\pi)^2}
\int^{\pi}_{-\pi}{d^2p_1\over(2\pi)^2}{d^2p_2\over(2\pi)^2}
{\tilde v}(\tilde{\bf k})\exp i\{{\bf p}_1\cdot{\bf X}
\nonumber\\
+{\bf p}_2\cdot{\bf Y}+
{\bf k}\cdot{\bf Z}+f({\bf p}_1,{\bf k})-f({\bf p}_2,{\bf k})\},
\label{vterm}
\end{eqnarray}
where $\tilde v({\bf k})=q^2 \exp(-{\bf k}^2/4\pi){\tilde V}({\bf k})$ 
and ${\tilde V}({\bf k})=2\pi/\vert{\bf k}\vert$ for ${\bf k}\neq0$ 
and ${\tilde V}(0)=0$,respectively. 
Here, $\tilde {\bf k}=(k_x/r,r k_y)$ and ${\bf p}_i$ is 
the lattice momentum in the Brillouin zone $\vert p_i\vert<\pi$. 
The phase function $f({\bf p},{\bf k})$ is defined by
\begin{equation}
f({\bf p},{\bf k})=-k_x(2p_y+k_y)/4\pi-\lambda({\bf p}+{\bf k})
+\lambda({\bf p}),
\end{equation}
where $\lambda({\bf p})$ is any real function satisfying the boundary 
condition 
$
e^{i\lambda({\bf p}+2\pi{\bf N})-i\lambda({\bf p})}
=(-1)^{N_x+N_y}e^{-iN_yp_x}.
$ 
Here, $N_x$ and $N_y$ are integers. 
Note that $f({\bf p},{\bf k})$ is written by using a line integral along 
a straight line as $\int_p^{p+k}({\bf A}({\bf p})-\partial\lambda({\bf p}))
\cdot d{\bf p}$ with ${\bf A}({\bf p})=(p_y/2\pi,0)$. 
${\bf A}({\bf p})$ represents a magnetic field in the momentum space 
and $\lambda({\bf p})$ corresponds to the gauge degree of freedom. 
The Hamiltonian $H$ is invariant under the transformation
\begin{equation}
b({\bf X})\rightarrow e^{i{\bf K}\cdot{\bf X}}b({\bf X}),
\label{KX}
\end{equation}
up to the gauge transformation\cite{m}. 
Eq.~(\ref{KX}) corresponds to the translation $\bf p\rightarrow \bf p+
\bf K$. 
This invariance is similar to the $\bf K$-invariance in the composite 
fermion model\cite{j}. 
In our case, on the other hand,  Eq.~(\ref{KX}) is equivalent to 
the magnetic translation in the momentum space. 

The interaction potential satisfies the following sum rules:
\begin{eqnarray} 
&\sum_{\bf Z} V({\bf X},{\bf Y},{\bf Z})=
{\tilde v}(2\pi{\hat{\bf X}})\delta^X_{-Y},
\label{sum}\\
&\sum_{\bf X} V({\bf X},{\bf Y}-{\bf X},{\bf Z})=
v({\hat{\bf Z}})\delta^Y_0,\nonumber
\end{eqnarray}
where $
v({\bf X})=\pi q^2 
e^{-(\pi/2){\bf X}^2}I_0((\pi/2){\bf X}^2)$, which is the Fourier 
transform of $\tilde v$, where $I_0$ is the modified Bessel function of 
the first kind and 
${\hat{\bf X}}=(rm,n/r)$ for ${\bf X}=(m,n)$. 
$\hat{\bf X}$ is a position of the Wannier basis in the real space. 

We introduce a mean field for the QHG state 
which has the translational invariance on the von Neumann lattice. 
Namely,
\begin{equation}
U_0({\bf X}-{\bf X}')=\langle b^\dagger({\bf X}')b({\bf X})\rangle,
\label{U0}
\end{equation}
and $U_0(0)=\nu$. 
If $U_0({\bf X})\neq0$ for ${\bf X}\neq0$, the symmetry under Eq.~
(\ref{KX}) is broken. 
Using the sum rules of Eq.~(\ref{sum}), the mean field Hamiltonian for 
$H_0$ becomes
\begin{eqnarray}
H_{\rm mf}&=&
\sum_{X,X'}\{-U_0({\bf X}-{\bf X}')
v({\hat{\bf X}}'-{\hat{\bf X}})b^\dagger({\bf X})b({\bf X}')\label{HFE}\\
&&+{1\over2}U_0({\bf X}-{\bf X}')
v({\hat{\bf X}}'-{\hat{\bf X}})U_0({\bf X}'-{\bf X})\}.\nonumber
\end{eqnarray}
The first term of the right-hand side 
is a kinetic energy dynamically induced from the Coulomb interaction. 
It is remarkable that the continuum system with a magnetic field 
is transformed to the lattice system without a magnetic field. 
In the composite fermion model, the external magnetic field is canceled 
by the Chern-Simons gauge field at $\nu=1/2$. 
In our model, on the other hand, 
the  magnetic field in the momentum space disappears for a mean field 
$U_0({\bf X}-{\bf X}')$ at arbitrary filling factor. 

By Fourier transforming Eq.~(\ref{HFE}), and using Eq.~(\ref{U0}), 
we obtain self-consistency equations for 
the one-particle kinetic energy $\varepsilon({\bf p})$ 
in the Hartree-Fock approximation as
\begin{eqnarray}
\varepsilon({\bf p},\nu)&=&-\int_{-\pi}^{\pi}{d^2 p_1\over (2\pi)^2}
{\tilde v}_{\rm f}({\bf p}_1-{\bf p})
\theta(\mu_0(\nu)-\varepsilon ({\bf p}_1,\nu)),\nonumber\\
\nu&=&\int_{-\pi}^{\pi}{d^2 p\over (2\pi)^2}
\theta(\mu_0(\nu)-\varepsilon({\bf p},\nu)),
\label{self}
\end{eqnarray}
where $\mu_0$ is the chemical potential for the mean field and 
$\tilde v_{\rm f}$ is defined by 
$
{\tilde v}_{\rm f}({\bf p})=\sum_{\bf N}\tilde v(\tilde{\bf p}+
2\pi\tilde{\bf N})$. 
Equation (\ref{self}) determines a self-consistent Fermi surface. 
Existence of the Fermi surface breaks $\bf K$-invariance. 
We obtain the solution that breaks $K_y$-invariance but preserves 
$K_x$-invariance. 
We use the lines $\vert p_y\vert=\pi\nu$ as 
the self-consistent Fermi surface which is a solution of Eq.~(\ref{self}) 
for an arbitrary $\nu$\cite{mx}. 
The one-particle states are filled in $\vert p_y\vert<\pi\nu$. 
Then the kinetic energy $\varepsilon({\bf p},\nu)$ becomes constant in the 
$p_x$-direction and is given by
\begin{eqnarray}
\varepsilon({\bf p},\nu)&=&\varepsilon(p_y,\nu)=
-\int_{-\pi\nu}^{\pi\nu}{dp\over 2\pi}
{\tilde v}_1(p_y-p),\\
{\tilde v}_1(p)&=&q^2r\sum_n e^{-r^2(p+2\pi n)^2/8\pi}
K_0(r^2(p+2\pi n)^2/8\pi),
\nonumber
\end{eqnarray}
where $K_0$ is the modified Bessel function of the second kind and 
the chemical potential $\mu_0$ is written as 
$\mu_0(\nu)=\varepsilon(\pi\nu,\nu)$. 
The total chemical potential $\mu$ is calculated by the total energy 
per particle $E(\nu)$ as $\mu=E(\nu)+\nu E'(\nu)$. 
The kinetic energy $\varepsilon(p_y,\nu)$ is obtained numerically and 
is shown in Fig.~1 for $\nu=1/4$, $1/2$, and $3/4$. 

Owing to the Coulomb interaction, the Fermi velocity 
$\partial\varepsilon/\partial p_y$ 
diverges logarithmically at the Fermi momentum $p_y=\pm\pi\nu$. 
Hence, there exists the Coulomb gap in the one-particle spectrum. 
This is consistent with the experiment\cite{mm} which suggests that there is 
no true metallic phase around $\nu=1/2$. 
Naively it seems that the divergence of the Fermi velocity makes the 
compressibility $\kappa$ vanishes because $\kappa$ is proportional to 
the inverse of the Fermi velocity in the ordinary electron gas. 
As is seen later, however,  
the $\nu$ dependence of $\varepsilon(p_y,\nu)$ contributes 
to $\kappa$ negatively. 
Furthermore, the distant background contribution to $\kappa$ is positive. 
Then these two contributions compete with each other and cause a transition.

First, we calculate the total energy of the QHG state. 
The energy per particle due to mean field Hamiltonian $H_{\rm mf}$ is 
given by
\begin{equation}
E_{\rm mf}(\nu)={1\over2\nu}\int_{-\pi\nu}^{\pi\nu}{dp_y\over2\pi}
\varepsilon(p_y,\nu).
\end{equation}
We calculate the fluctuation around the mean field and find 
the Hartree term contributes to the energy per particle in the lowest 
order as
\begin{eqnarray}
E_{\rm h}(\nu)&=&{1\over 2\nu}\sum_X U_0({\bf X}){\tilde v}(2\pi{\hat
{\bf X}})
U_0(-{\bf X})\\
&=&q^2\sum_{n=1}^{\infty}
{re^{-\pi(n/r)^2}\sin^2(\pi\nu n)\over\pi^2\nu n^3}. 
\nonumber
\end{eqnarray}
We approximate the total energy per particle $E(\nu)$ as 
\begin{eqnarray}
E(\nu)&=&E_0(\nu)+2\pi q^2\nu d,\label{total}\\
E_0(\nu)&=&E_{\rm mf}(\nu)+E_{\rm h}(\nu).
\nonumber
\end{eqnarray}
$E_0(1/2)$ becomes minimum at $r=1.64$, which is used in this letter. 
The minimum value of $E_0(1/2)$ is $-1.086 q^2$ 
$(=-0.433 q^2/l_B$, $l_B=\sqrt{\hbar/{\rm e}B}$). 
This value is slightly higher than values of the charge density wave 
(CDW) calculation\cite{CDW} which is shown in Fig.~2 by a dot. 
In the CDW, the higher order correction is small because of the energy gap. 
In the QHG, however, the higher order correction might be large compared with 
the CDW. 
The value of $E_{\rm h}$ is positive and is smaller than one tenth of 
the magnitude of $E_{\rm mf}$. 
Although $E_{\rm h}$ has a small contribution to the energy, 
this term is necessary for $E_0(1/2)$ to have the minimum 
as a function of $r$ because $E_{\rm mf}$ and 
$E_{\rm h}$ are monotonically decreasing and increasing function of $r$, 
respectively, at fixed filling factor. 
In Fig.~2, $E(\nu)$ is plotted for $d=0$, $0.1$, and $0.2$. 
At $\nu=1/3$, the value of Laughlin's incompressible state is shown in Fig.~2 
by a cross. 
The energy of the QHG for $d=0$ is substantially 
higher than the Laughlin's value. 
Therefore the incompressible state is preferable to the QHG state. 

Second, we calculate the pressure of the QHG state. 
The pressure $P$ is defined by the derivative of the total energy per 
particle as $P=\nu^2 E'(\nu)$. 
In Fig.~3 the pressure is plotted for $d=0$, $0.1$, and $0.2$. 
Using the particle-hole transformation $b^\dagger\leftrightarrow b$, 
it can be shown that $E'(1/2)$ obeys the relation 
$E'(1/2)=-2(E(1/2)-E(1))$, 
where $E(1)=-\pi q^2/2+2\pi q^2 d$. 
Therefore the pressure at the half-filling becomes negative 
below a critical value $d=d_{c1}$, which is determined by $E(1/2)=E(1)$ 
as $d_{c1}=(E_0(1/2)+\pi q^2/2)/\pi q^2=0.154$ 
in the unit of $a$. 
For $\nu=1/4$ and $3/4$, the critical values are numerically obtained 
as $0.253$ and $0.138$ respectively. 
As is seen in Fig.~2, $E(\nu)$ is a decreasing function for $d=0$. 
This behavior was verified by the exact diagonalization of the 
small system\cite{n}. 
Thus, if $E(\nu)$ is a smooth function, the pressure is negative for 
$d=0$. 
The negative pressure does not always mean the instability 
of the gas state. 
For the stability, the positive compressibility is 
necessary. 

Finally, let us calculate the compressibility of the QHG state. 
The compressibility $\kappa$ is defined by the derivative of the pressure 
$P(\nu)$ as $1/\kappa=\nu P'(\nu)$. 
In calculating $\kappa$ it is rather convenient to use the relation
$1/\kappa=\nu^2\mu'(\nu)$. 
Using Eq.~(\ref{total}) we obtain 
\begin{eqnarray}
\mu'(\nu)&=&-{\tilde v}_1(2\pi\nu)+\sum_{n=1}^{\infty}2q^2re^{-\pi(n/r)^2}
\cos(2\pi\nu n)/n\nonumber\\
&&+4\pi q^2 d.
\label{munu}
\end{eqnarray}
The first term of the r.h.s. comes from $\mu_0$ and the second term 
comes from $E_{\rm h}$. 
In Fig.~4, $1/\mu'(\nu)$, which is proportional to $\kappa$, 
is plotted for $d=0$, $0.1$, and $0.2$. 
The positive $\kappa$ region appears symmetrically with respect to 
$\nu=1/2$ for $d>d_{c2}$. 
The critical value $d_{c2}$ at $\nu=1/2$ is determined by 
$\mu'(1/2)=0$ as $d_{c2}=0.115$ in the unit of $a$. 
For $\nu=1/4$ and $3/4$, the critical values are also obtained 
as $0.151$. 

The negative compressibility has been observed by using the electric field 
penetrating technique\cite{o}. 
This method makes it possible to measure the compressibility of the 
two-dimensional electron gas discriminating 
the background charge contribution. 
The observed negative compressibility may correspond
 to the Eq.~(\ref{munu}) for $d=0$. 
Near $\nu=0$ and $1$ the compressibility is always negative in Fig.~4. 
A transition between the QHG and the Wigner crystal is expected when the 
$\kappa$ diverges. 
Note that $d_{c1}>d_{c2}$ at $\nu=1/4$ and $1/2$, and 
$d_{c1}<d_{c2}$ at $\nu=3/4$. 
At $B\approx10$ T, $d_{c1}$ and $d_{c2}$ are on the order of 1 nm. 
By varying $B$, $d_{c1}$ and $d_{c2}$ vary as $\propto B^{-1/2}$. 
In the high-mobility GaAs-AlGaAs heterostructure, $d>$10 nm, 
the QHG state becomes stable. 
Changing a magnetic field or a width of the spacer, 
the transition could be observed in experiments. 
The QHG state might be changed into the incompressible liquid state 
or the insulator phase. 

In the composite fermion model, the Fermi wave number $k_F$ at 
$\nu=1/2$ is given by $k_F=\sqrt{4\pi n_e}$, $n_e$ is the 
density of the electrons, which was verified by experiments
\cite{c,d,e}. 
In our model, the Fermi wave number $p_F$ on the von Neumann lattice 
is given by $p_F=\pi\nu$, which corresponds to the 
the Fermi wave number $k_F$ on the real space given by 
$k_F=\pi\nu r/a=r\pi\sqrt{\nu n_e}$. 
For $r=1.64$ and $\nu=1/2$, $r\pi\sqrt{\nu}=3.64$, which is very close 
to the value $\sqrt{4\pi}=3.54$. 

In summary, we study the QHG state at arbitrary $\nu$ 
using the von Neumann lattice. 
The QHG state has a Fermi surface and is stabilized by the spacer 
around $\nu=1/2$ for $d>d_{c2}$ 
and a transition occurs for $d<d_{c2}$ in the quantum Hall system 
without disorder. 

Authors thank Prof. P. Wiegmann for useful discussions. 
This work was partially supported by the special Grant-in-Aid
for Promotion of Education and Science in Hokkaido University
provided by the Ministry of Education, Science, Sports and Culture,
the Grant-in-Aid for Scientific Research on Priority Area (Physics of CP 
violation) (Grant No. 10140201), and the
Grant-in-Aid for International Scientific Research (Joint Research
 Grant No. 10044043) from
the Ministry of Education, Science, Sports and Culture, Japan.

%%%%%%%%%%%%%%%%%%%%%%%%%%%%%%%%%%%%%%%%%%%%%%%%%%%%%%%%%%%%%%%%%%%%%%%%
\begin{figure}
\centerline{
\epsfysize=1.7in\epsffile{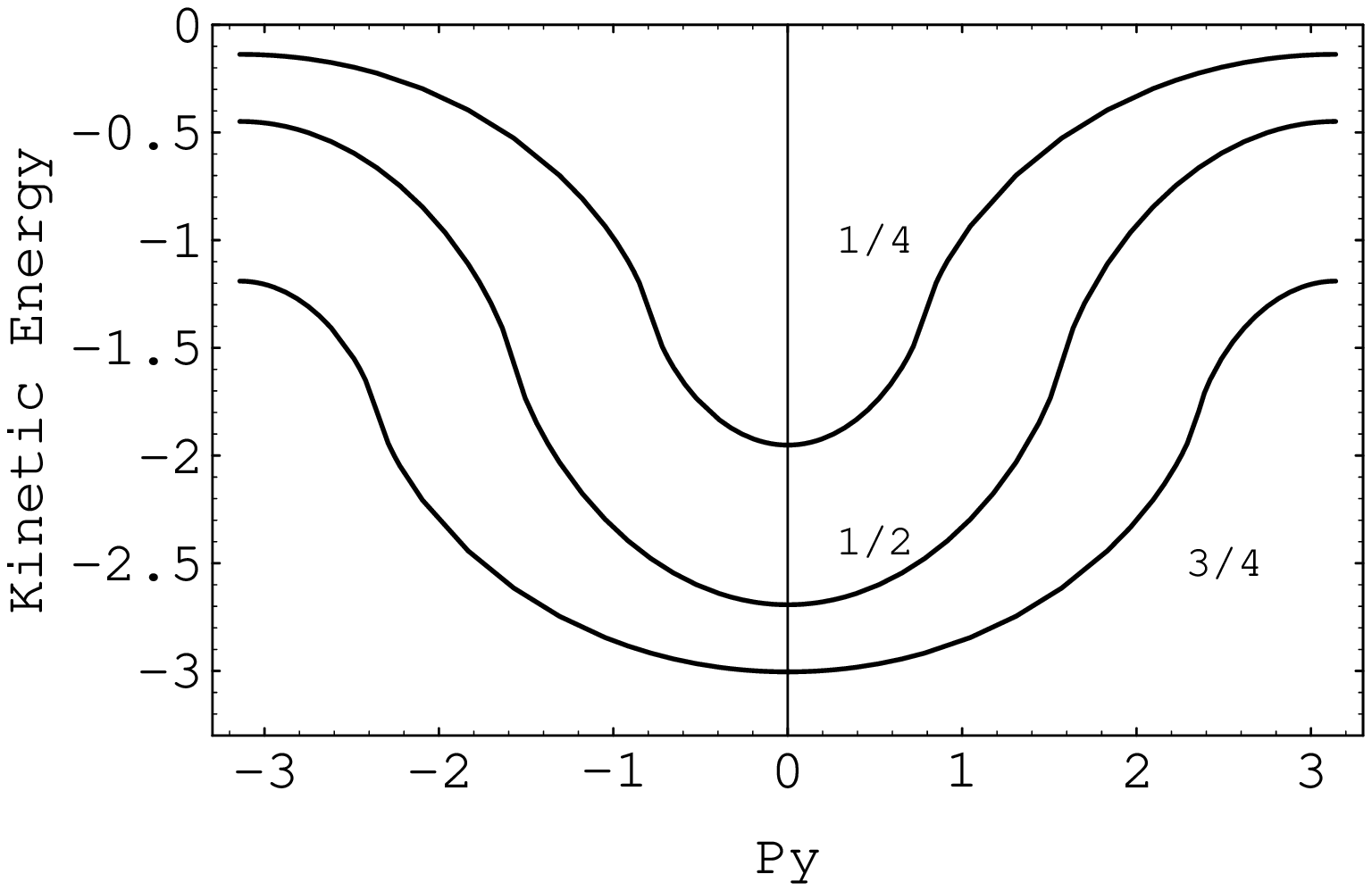}}
Fig~1. 
Kinetic energy $\varepsilon(p_y,\nu)$ for $\nu=1/4$, 1/2, and 3/4. 
The unit of the energy is $q^2/a$. 
The gradient of $\varepsilon$ diverges at $p_y=\pm\pi\nu$.
\end{figure}
%%%%%%%%%%%%%%%%%%%%%%%%%%%%%%%%%%%%%%%%%%%%%%%%%%%%%%%%%%%%%%%%%%%%%%%%%
\begin{figure}
\centerline{
\epsfxsize=2.5in\epsffile{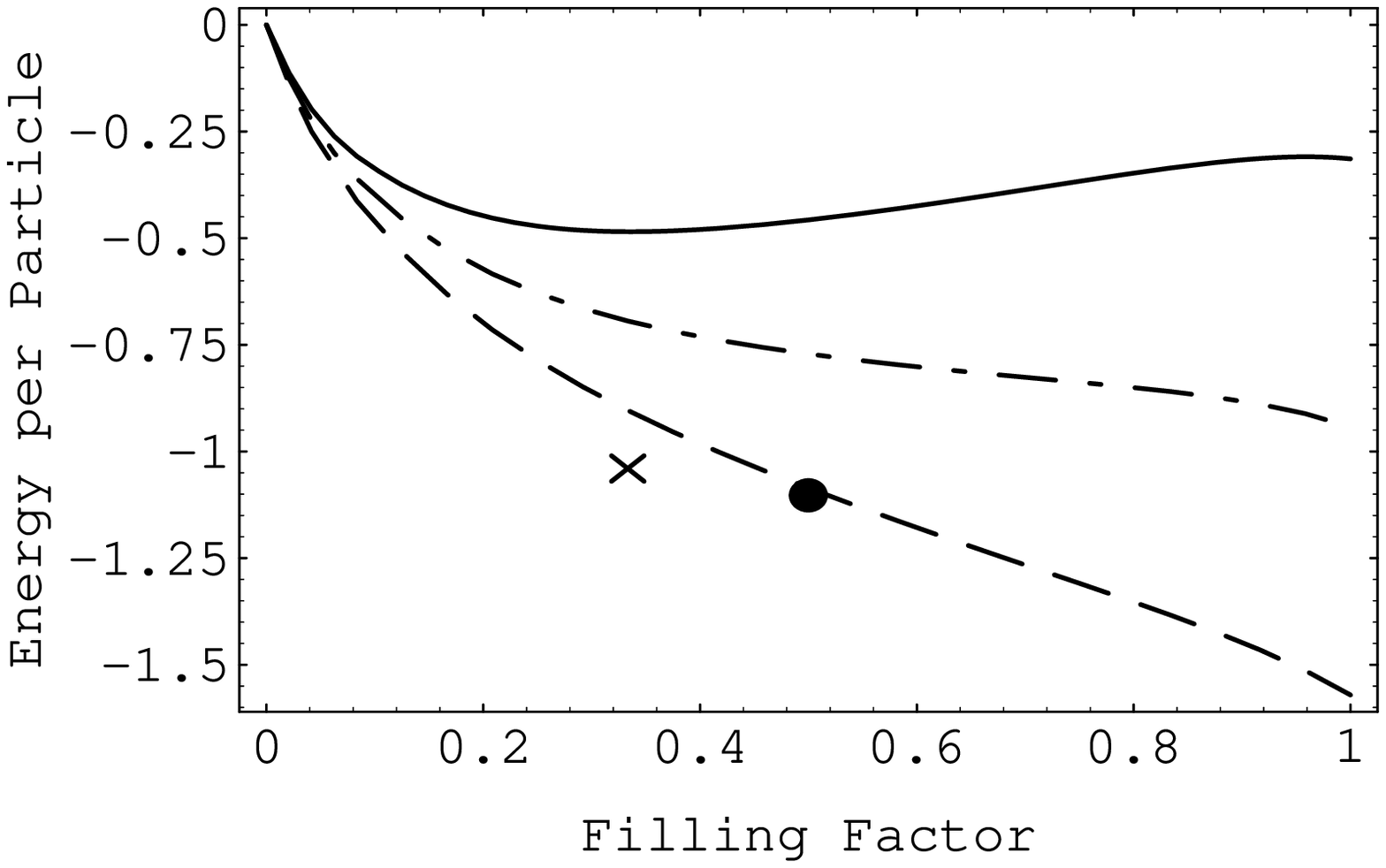}}
Fig.~2. 
Energy per particle $E(\nu)$ for $d=0$ (dashed line), 0.1 (
dash-dotted line), and 0.2 (solid line). 
The unit of the energy is $q^2/a$. 
The cross ($\times$) shows Laughlin's value at $\nu=1/3$ and dot ($\bullet$) 
shows CDW value at $\nu=1/2$. 
\end{figure}
%%%%%%%%%%%%%%%%%%%%%%%%%%%%%%%%%%%%%%%%%%%%%%%%%%%%%%%%%%%%%%%%%%%%%%%%%
\begin{figure}
\centerline{
\epsfysize=2in\epsffile{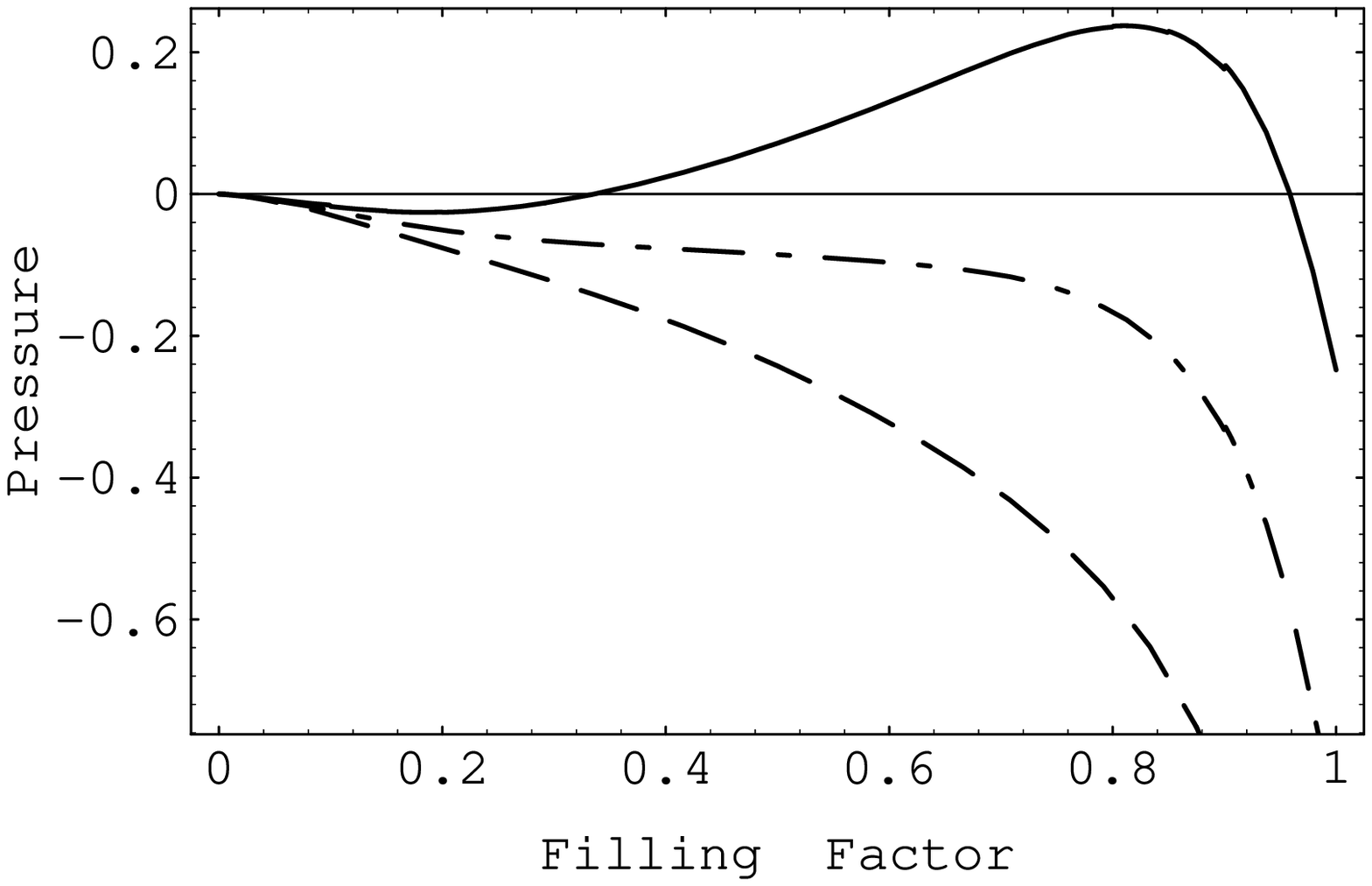}}
Fig~3. 
Pressure $P(\nu)$ for $d=0$ (dashed line), 
0.1 (dash-dotted line), and 0.2 (solid line). 
The unit of the pressure is $q^2/a^3$.
\end{figure}
%%%%%%%%%%%%%%%%%%%%%%%%%%%%%%%%%%%%%%%%%%%%%%%%%%%%%%%%%%%%%%%%%%%%%%
\begin{figure}
\centerline{
\epsfysize=1.9in\epsffile{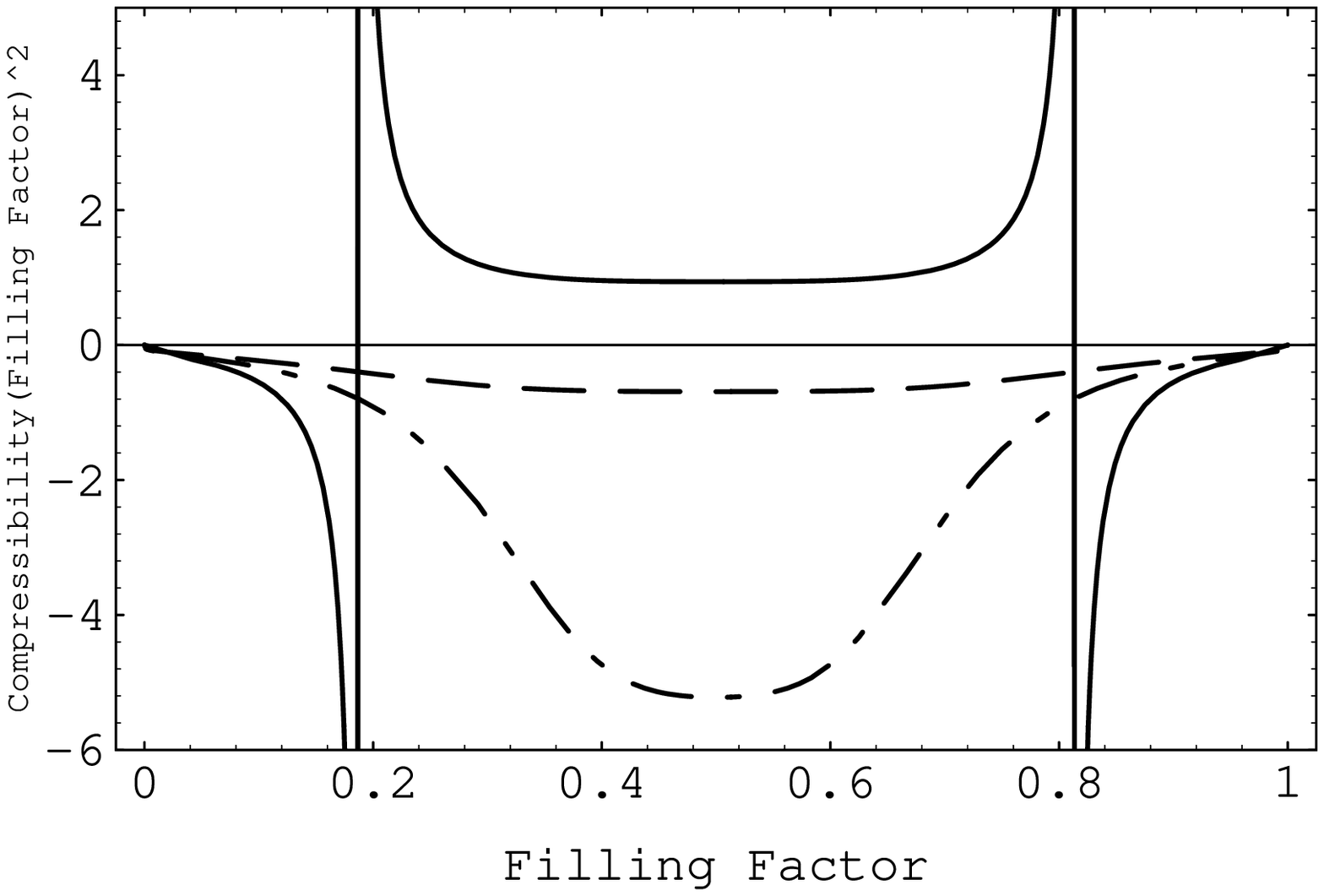}}
Fig.~4. 
Compressibility $\kappa(\nu)$ times $\nu^2$ 
for $d=0$ (dashed line), 
0.1 (dash-dotted line), and 0.2 (solid line).  
The unit of the $\kappa$ is $a^3/q^2$.
\end{figure}
%%%%%%%%%%%%%%%%%%%%%%%%%%%%%%%%%%%%%%%%%%%%%%%%%%%%%%%%%%%%%%%%%%%%%%

\end{multicols}

\begin{references}
\bibitem{mx}
At $\nu=1/2$, the Fermi surface $\cos p_x+\cos p_y=0$ also becomes 
a self-consistent solution. 
We find that its energy is slightly higher than the 
one in the text.
\end{references}
\end{document}